\documentclass[aps,prd,twocolumn,showpacs,eqsecnum,A4,amsmath,amssymb,nofootinbib]{revtex4-1}

\usepackage{amsfonts}
\usepackage{amsmath}
\usepackage{amssymb}
\usepackage{graphicx}
\usepackage{color}
\usepackage{appendix}

\newcommand{\dd}{{\rm{d}}} 

\newcommand{\im}{\mathrm{i}}

\begin{document}

\title{Kerr black hole in a uniform Bertotti-Robinson magnetic field: An exact solution}

\author{Ji\v{r}\'i Podolsk\'{y}}
\email{jiri.podolsky@mff.cuni.cz}
\affiliation{Charles University, Faculty of Mathematics and Physics,
Institute of Theoretical Physics,
V~Hole\v{s}ovi\v{c}k\'ach 2, 18000 Prague 8, Czechia}

\author{Hryhorii Ovcharenko}
\email{hryhorii.ovcharenko@matfyz.cuni.cz}
\affiliation{Charles University, Faculty of Mathematics and Physics,
Institute of Theoretical Physics,
V~Hole\v{s}ovi\v{c}k\'ach 2, 18000 Prague 8, Czechia}

\begin{abstract}
A new class of exact spacetimes in Einstein's gravity, which are Kerr black holes immersed in an external magnetic (or electric) field that is asymptotically uniform and oriented along the rotational axis, is presented. These are axisymmetric stationary solutions to the Einstein-Maxwell equations such that (unlike in the Pleba\'nski-Demia\'nski spacetime) the null directions of the Faraday tensor are not aligned with neither of the two principal null directions of the Weyl tensor of algebraic type~D (unlike the Kerr-Melvin spacetime). Three physical parameters are the black hole mass~$m$, its rotation $a$, and the external field value $B$. For vanishing $B$ the metric directly reduces to standard Boyer-Lindquist form of the Kerr black hole, while for zero $m$ we recover conformally flat Bertotti-Robinson universe with a uniform Maxwell field. For zero $a$ the spacetime is contained in the Van den Bergh-Carminati solutions which can be understood as the Schwarzschild black hole in a magnetic field. Our family of black holes with non-aligned Maxwell hair --- that can be called the Kerr-Bertotti-Robinson (Kerr-BR) black holes --- may find application in various studies ranging from mathematical relativity to relativistic astrophysics.
\end{abstract}

\date{\today}



\maketitle


\newpage

\section{Introduction}

Black holes, regions with extremely strong gravity, are remarkable theoretical predictions of Einstein's general relativity and fascinating objects in our Universe. Direct observational evidences of their existence were recently provided by detections of gravitational waves from binary black hole coalescences \cite{LIGO-Virgo:2016}, and by the first images of a shadow of supermassive black holes in M87* \cite{EHT:2019} and Sgr~A*, the center our Galaxy.

The most important \emph{exact} models of black holes are the Schwarzschild (1916) and the Kerr (1963) solutions. They are \emph{unique} spherically/axially symmetric stationary vacuum spacetimes that are asymptotically flat \cite{Stephani:2003tm, GriffithsPodolsky:2009}. These textbook metrics have been widely employed for investigation of their surprising mathematical structure and, above all, of (astro)physical effects in the vicinity of the static/rotating black holes, such as influence on matter and fields, thermodynamics, or quantum phenomena.

Their charged versions exist. The Reissner-Nordstr\"{o}m (1916, 1918) and Kerr-Newman (1965) black holes
are solutions to the Einstein-Maxwell equations such that the algebraic structures of the gravitational and electromagnetic fields are \emph{fully aligned} \cite{Stephani:2003tm, GriffithsPodolsky:2009}. Generalizations include a cosmological constant, NUT parameter, or acceleration. Interestingly, all these exact spacetimes belong to a big class such that the Weyl tensor is of algebraic type~D and the two null directions of (non-null) Faraday tensor are both aligned with (double-degenerate) principal null directions (PNDs) of the Weyl tensor \cite{Debever1971, Plebanski1976, Debever1984}. Pleba\'nski and Demia\'nski in their seminal work \cite{Plebanski1976} found a compact representation of this class, the metric forms more suitable for physical studies of these type D black holes were obtained in \cite{GRIFFITHS2006,Griffiths2005,PodolskyGriffiths:2006} and \cite{Podolsk2021,Podolsk2023}. A new description of the whole family was found recently  \cite{Astorino:2024b,Ovcharenko2024, Ovcharenko2025}.

Let us remark that these spacetimes in general possess conical singularities (strings or struts) related to acceleration and NUT twist. For asymptotically flat Kerr-Newman black holes the axis of symmetry is regular.

Another generalization of the Kerr-Newman family was found by adding an \emph{external electromagnetic field}. This was motivated astrophysically  because black holes in the Universe are usually immersed in a magnetic field, providing a basis for the explanation of the high energy activity of galactic nuclei and quasars. Such exact solutions were first found by Ernst and Wild \cite{Ernst1976_2, Ernst1976_3} employing the Harrison transformation.
These were nicknamed Schwarzschild-Melvin, Kerr-Melvin, and Kerr-Newman-Melvin spacetimes because they can be understood as the corresponding black holes immersed in an external magnetic field of the \emph{Bonnor-Melvin universe} \cite{Bonor:1954, Melvin:1964}, see \cite{GriffithsPodolsky:2009} for more details. Further interesting extensions were found, most recently in \cite{DiPinto2025}.

These black holes have been widely used for the analysis of various physical phenomena, see \cite{Galtsov1989, Gibbons2014} for reviews of the main results. In particular, interaction between the external field and the black hole was investigated, leading to the discovery of the so-called \emph{Meissner effect}, namely that black holes expel magnetic field away from their horizon if they become extremal, see e.g. \cite{Wald1974,BicakDvorak1980,Bicak2015,GurlebeckScholtz:2018}.

The Melvin-like black hole spacetimes, however, have also some drawbacks preventing them to be considered as fully realistic global models. The magnetic field decreases far away from the black hole, but geodesics cannot escape to infinity. They can even be chaotic \cite{KarasVokrouhlicky:1992}. Ergoregions extend to infinity \cite{Gibbons2013}. Moreover, the spacetimes are  of algebraic type~I, see \cite{Pravda2005}, suggesting that such ``standing''  black holes can be seen as radiative.

\newpage
Our paper aims to overcome these problems. In fact, we will show that the new class of spacetimes is of algebraic type~D, has bounded ergoregions, and the field is asymptotically finite and uniform. The test particles can, in principle, escape to infinity. Such properties~\hbox{allow} us~to consider this class as a more appropriate model of a black hole immersed in an external electromagnetic field.

In Sec.~\ref{section-e=0=g} the new class of solutions is presented, and in Sec.~\ref{sec:physics} we perform its fundamental physical analysis.

\section{The Kerr-BR metric and the electromagnetic field}
\label{section-e=0=g}

Our novel metric has a compact explicit form
\begin{align}
    \dd s^2=\dfrac{1}{\Omega^2}\Big[&-\dfrac{{Q}}{\rho^2}
    \big(\dd t-a\sin^2\theta\,\dd\varphi\big)^2
    +\dfrac{\rho^2}{{Q}}\,\dd r^2
    +\dfrac{\rho^2}{{P}}\,\dd\theta^2\nonumber\\
    &+\dfrac{{P}}{\rho^2}\sin^2\theta\,
    \big(a\,\dd t-(r^2+a^2)\,\dd\varphi\big)^2\,\Big],\label{Kerr-BR}
\end{align}
where the metric functions are
\begin{align}
\rho^2   & = r^2+a^2\cos^2\theta\,, \label{rho2}\\[0mm]
{P} & = 1 + B^2 \Big(m^2\,\dfrac{I_2}{I_1^2} - a^2 \Big)\cos^2\theta\,,  \label{tilde_P}\\
{Q} & = \,\big(1+B^2r^2\big)\, \Delta\,,\label{math-Q}\\[2mm]
\Omega^2 & = \,\big(1+B^2r^2\big) - B^2 \Delta \cos^2\theta\,,\label{Omega}\\
\Delta &= \Big(1-B^2m^2\,\dfrac{I_2}{I_1^2}\Big) r^2-2m\,\dfrac{I_2}{I_1}\,r + a^2
\,,\label{Delta}
\end{align}
with
\begin{align}
    I_1 = 1-\tfrac{1}{2} B^2a^2  \,,\qquad
    I_2 = 1-B^2a^2   \,.\label{I1I2}
\end{align}
The electromagnetic field is given by the complex 1-form potential (its real counterpart is ${\mathbf{A}^{\rm real} \equiv  2\,{\rm Re}\,\mathbf{A} }$)
\begin{align}
    \mathbf{A} = \dfrac{{\rm e}^{\im\,\gamma}}{2B}\,&\Big[\,
    \Omega_{,r}\,\dfrac{ a\,\dd t-({r}^{\,2}+a^2)\,\dd\varphi }{r+\im\,a\cos\theta}  \nonumber\\
    &+\dfrac{\im\,\Omega_{,\theta}}{\sin\theta}\,\dfrac{ \dd t - a\sin^2\theta\,\dd\varphi}
       {r+\im\,a\cos\theta} + (\Omega-1)\,\dd\varphi\, \Big] ,\label{A-7}
\end{align}
which vanishes for ${B=0}$. Systematic derivation of this spacetime, with a generalization to include charge and acceleration, is given in \cite{OP-prepar2}. In the null complex tetrad
\begin{align}
    \mathbf{k}&=\dfrac{\Omega}{\rho\,\sqrt{2{Q}}}\,
       \big((r^2+a^2)\,\partial_{t}+a\,\partial_{\varphi}+{Q}\,\partial_{r}\big),\nonumber\\
    \mathbf{l}&=\dfrac{\Omega}{\rho\,\sqrt{2{Q}}}\,
       \big((r^2+a^2)\,\partial_{t}+a\,\partial_{\varphi}-{Q}\,\partial_{r}),\label{null_tetr_trans}\\
    \mathbf{m}&=\dfrac{-\Omega}{\rho\,\sqrt{2{P}}\sin\theta}\,
       \big(a\,\sin^2\theta\,\partial_{t}+\partial_{\varphi}+\im\,{P}\sin\theta \,\partial_{\theta}\big),\nonumber
\end{align}
the only Weyl curvature component is given by the scalar
\begin{align}\label{Psi2}
    \Psi_2 = - \frac{m}{(r+\im\, a \cos\theta)^3}\,&
    \bigg[\,\big(1 - \im \,B^2 a\,r \cos\theta\big) I_1 I_2 \nonumber\\
          &+B^2 \,\dfrac{m\, r^2\cos^2\theta}{r-\im\, a \cos\theta}\,\bigg]
     \,\dfrac{\Omega^2}{I_1^2}\,,
\end{align}
so that the spacetime is of algebraic type~D.

The \emph{nonaligned parts} of electromagnetic field \cite{Stephani:2003tm, GriffithsPodolsky:2009} are
\begin{align}
    \Phi_0=\Phi_2 = B\,{\rm e}^{\im\,\gamma} \,\dfrac{1}{2\Omega}\,
    \dfrac{\sqrt{{P}{Q}}\,\sin\theta }{r+\im\,a\cos\theta}\,.
        \label{Phi0,1=0,e=0=g}
\end{align}
Interestingly, they vanish on the horizon (where ${{Q}=0}$) and also along the axis of symmetry (at ${\theta=0, \pi}$).
The \emph{aligned part} is more involved, namely
\begin{align}
    \Phi_1 =  B\,{\rm e}^{\im\,\gamma} \,\,
    \dfrac{B_0+B_1\,r+B_2\,r^2+B_3\,r^3}{2 \Omega\, I_1^3\,(r+\im\,a\cos\theta)^2}\,,
     \label{Phi1,l=0,e=0=g}
\end{align}
where $B_i(\theta)$ are the $r$-independent functions
\begin{align}\label{Bi}
B_0 &= a\, I_1^2\cos\theta\,\big[ m\,I_2 \cos\theta
   - \im\,a\,I_1 (1-B^2 a^2 \cos^2\theta) \big],\nonumber\\
B_1 &= a\, I_1 \big[\!-I_1^2(1+\cos^2\theta) \nonumber\\
      &\hspace{6mm} + B^2 \cos^2\theta \,( \,m^2 I_2
     +  2 a^2I_1^2 - \im\,3 a m I_1I_2 \cos\theta)\big],\nonumber\\
B_2 &= I_1 \big[-3 B^2 am\,I_1 I_2 \cos^2\theta - \im\,D\,B^2a^2 \\
 &\hspace{6mm}  + \im\,\cos\theta\,\big(I_1^2-B^2 m^2I_2\,(1-2I_2\cos^2\theta) \big)
    \big], \nonumber\\
B_3 &= -B^2 \big[ \,a\, I_1 (I_1^2\sin^2\theta+B^2m^2I_2\cos^2\theta) -\im\,D\,m\, I_2 \,\big] ,\nonumber
\end{align}
${D= \cos\theta \,[\,I_1^2(2-\cos^2\theta) + B^2 m^2 I_2\cos^2\theta\,]}$,
and $\gamma$ is the duality rotation parameter. For ${m=0}$, purely magnetic field is given by ${\gamma=0}$, while purely electric field is given by ${\gamma=\tfrac{\pi}{2}}$. Clearly, the \emph{Maxwell field vanishes if and only~if ${B=0}$}. The corresponding invariant
\begin{align}
\tfrac{1}{16}\,F^*_{\,\mu\nu}&\,F^{*\,\mu\nu} = \Phi_0 \Phi_2-\Phi_1^2 \label{EM-invariant}
\end{align}
is non-zero, so that the electromagnetic field is non-null.

To summarize, the metric \eqref{Kerr-BR} satisfies the Einstein-Maxwell equations with  ${\mathbf{A}^{\rm real} \equiv 2\,{\rm Re}\,\mathbf{A} }$ given by \eqref{A-7}. It is of Weyl type~D (unlike the Kerr-Melvin type~I solution) with a non-aligned (and non-null) electromagnetic field without sources (unlike the Kerr-Newman solution which has an aligned Maxwell field generated by charges of the black hole). It is thus a new interesting class of exact spacetimes with three parameters, namely $m$, $a$, and $B$. We will now show that these can be physically interpreted as the \emph{mass} of the black hole, its \emph{rotation}, and \emph{value of the external uniform magnetic field}, respectively. Such an interpretation follows from the fact that the metric \eqref{Kerr-BR} reduces to usual forms of the rotating Kerr black hole and to the Bertotti-Robinson universe with a uniform Maxwell field when ${B=0}$ and ${m=0}$, respectively.

\subsection{The case ${B=0}$ is Kerr}
\label{sec:Kerr}

For vanishing Maxwell field, that is for ${B=0}$, the metric functions \eqref{rho2}--\eqref{Delta} simplify to ${\Omega^2 = 1}$, ${{P} = 1}$, ${\rho^2 = r^2+a^2\cos^2\theta}$, ${ {Q} = \Delta = r^2-2m\,r + a^2}$, so that the metric \eqref{Kerr-BR} becomes
\begin{align}
    \dd s^2=&-\dfrac{\Delta}{\rho^2}
    \Big(\dd t-a\sin^2\theta\,\dd\varphi\Big)^2
    +\dfrac{\rho^2}{\Delta}\,\dd r^2
    +\rho^2\,\dd\theta^2\nonumber\\
    &+\dfrac{\sin^2\theta}{\rho^2}\,
    \Big(a\,\dd t-(r^2+a^2)\,\dd\varphi\Big)^2\,.\label{Kerr}
\end{align}
This is exactly the famous form of the \emph{Kerr black hole} in Boyer-Lindquist (spheroidal) coordinates. It depends on two physical parameters, namely the mass $m$ and the rotation parameter~$a$ related to the angular momentum.

\subsection{The case ${m=0}$ is Bertotti-Robinson}
\label{sec:BR}

For ${m=0}$ the spacetime is conformally flat (${\Psi_2=0}$) with a non-null (and source-free) electromagnetic field. According to the famous uniqueness theorem (see Sec.~7.1 in \cite{GriffithsPodolsky:2009}) it must be the \emph{Bertotti-Robinson spacetime}. (If ${B=0}$ it is just Minkowski.)

This fact can be shown explicitly. The metric functions reduce to ${{P}=1-B^2 a^2\cos^2\theta}$, ${{Q}=(1+B^2r^2)(r^2 + a^2)}$, ${\Omega^2=P+B^2r^2\sin^2\theta}$, ${\rho^2=r^2+a^2\cos^2\theta}$, and a coordinate transformation ${r, \theta, t, \varphi \mapsto R, \Theta, \tau, \phi}$
\begin{align}
\frac{R^2}{e^2}&= P\,\frac{1+B^2r^2}{\Omega^2}-1\,, \qquad
e\,\sin \Theta = \frac{\sqrt{r^2+a^2}}{A \,\Omega} \sin\theta \,,\nonumber\\
\tau &= A^{-1}t\,, \qquad
\phi = A^2\varphi +B^2a\,t\,,\label{BR-trans}
\end{align}
where ${A=\sqrt{1-B^2a^2}}$ and ${e=(AB)^{-1}}$, puts \eqref{Kerr-BR} to the usual form of the Bertotti-Robinson spacetime
\begin{align}
    \dd s^2= & -(1+R^2/e^2)\,\dd \tau^2 + (1+R^2/e^2)^{-1}\,\dd R^2 \nonumber\\[2mm]
           & +e^2 (\dd \Theta^2+\sin^2\Theta \,\dd\phi^2)\,,
           \label{metric-BR}
\end{align}
see Eq.~(7.4) in~\cite{GriffithsPodolsky:2009}. Recall that it is a unique conformally flat homogeneous universe without singularities, filled with a uniform electromagnetic field. It has the direct-product $\mathrm{AdS}_2\times \mathrm{S}^2$ geometry\footnote{Asymptotic topology of a general Kerr-BR solution could be different. Also, ${r\to \infty}$ is not the conformal infinity because (\ref{Psi2}) does not vanish there. The proper conformal infinity and causal structure will be investigated elsewhere.} of 2-dim anti-de~Sitter spacetime and a \hbox{2-sphere} of constant radius~$e$.
Actually, \eqref{EM-invariant} for ${m=0}$  gives the constant Bertotti-Robinson value
\begin{align}
F^*_{\,\mu\nu}\,F^{*\,\mu\nu} = 4 {\rm e}^{\im\,2\gamma}\, B^2(1-B^2a^2) = (2 {\rm e}^{\im\,\gamma}/e)^2\,, \label{invariant-m=0}
\end{align}
because \eqref{Phi1,l=0,e=0=g} reduces to much simpler expression ${ \Phi_1 =  B\,{\rm e}^{\im\,\gamma} \,
    [\,\im (1-B^2a^2)\cos\theta-a\,\Omega^2\,]/[\,2 \Omega\,(r+\im\,a\cos\theta)\,]}$.

Such electromagnetic field can be put into a \emph{canonical non-null form}. The Bertotti-Robinson spacetime is  conformally flat, so that there is no PND of the Weyl tensor and we are allowed to perform a suitable null rotations with fixed~$\mathbf{k}$, and then fixed~$\mathbf{l}$. We obtain ${\Phi_1'' = \tfrac{\im}{2} AB\,{\rm e}^{\im\,\gamma}}$, ${\Phi_2'' = 0 = \Phi_0''}$, in full agreement with \eqref{invariant-m=0}.

\subsection{The case ${a=0}$ is Schwarzschild-BR}
\label{sec:SchwBR}

The metric \eqref{Kerr-BR} simplifies to
\begin{align}
    \dd s^2&=\dfrac{1}{\Omega^2}\Big[-{\cal{Q}}\,\dd t^2
           +\dfrac{\dd r^2}{{\cal{Q}}}
           +r^2\Big(\dfrac{\dd \theta^2}{{P}}+{P}\sin^2\theta \,\dd\phi^2 \Big)\Big],
           \nonumber\\[2mm]
{P}&=1+B^2m^2 \cos^2\theta\,, \nonumber\\
{\cal{Q}}&= \,\big(1+B^2r^2\big)\Big( 1 - B^2m^2 - \frac{2m}{r} \,\Big) ,
     \label{metric-Schw-BR}\\
\Omega^2& = 1 + B^2 \Big[\,r^2 \sin^2\theta + \big( 2 m\,r  + B^2m^2 r^2 \big)\cos^2\theta \Big].\nonumber
\end{align}
For ${B=0}$ we recover spherical form of the Schwarzschild solution.
The Weyl scalar \eqref{Psi2} reduces~to
\begin{align}\label{Psi2-a=0}
    \Psi_2 = - \frac{m}{r^3}\,
    \big(1 + B^2 m\, r\,\cos^2\theta\,\big)
     \,\Omega^2\,,
\end{align}
so that the curvature singularity is located at ${r=0}$, as for the Schwarzschild black hole. Far away from the black hole (for large~$r$) the curvature is finite, depending on~$\theta$. In the equatorial plane ${\theta=\frac{\pi}{2}}$ the spacetime is asymptotically flat because ${\Psi_2 = -m\,(1+B^2r^2)/r^3 \to 0}$.

The Maxwell field is obtained from \eqref{Phi0,1=0,e=0=g}--\eqref{Bi} as
\begin{align}
\Phi_0  &=\Phi_2= B\,{\rm e}^{\im\,\gamma} \,\,\frac{1}{2 \Omega}\,\sqrt{{{P}{\cal{Q}}}}\,\sin\theta \,, \nonumber\\
\Phi_1 &=  B\,{\rm e}^{\im\,\gamma} \,\,\frac{\im}{2 \Omega}\,(D_0+D_1 r) \cos\theta\,,
\label{Phi01,1=0,e=0=g}
\end{align}
where the coefficients are ${D_0 = 1 - B^2 m^2 (1-2\cos^2\theta)}$ and
${D_1 = B^2 m \,[\, 2 + (B^2 m^2 - 1) \cos^2\theta\, ]}$.

It can be shown \cite{OP-prepar1} that this type D solution with non-aligned Maxwell field is a subcase of the ${h=0}$ case of \cite{VandenBergh2020}, namely Eq.~(85) and Eqs.~(101)--(103) therein, expressed here in more convenient coordinates. This solution seems to be different from the solution presented in \cite{AlexeevGarcia1996} because it is of type~I \cite{OrtaggioAstorino:2018}.

\section{Physical analysis}
\label{sec:physics}

The novel family of black holes with non-aligned ``Maxwell hair'' can thus be justifiably called the \emph{Kerr-Bertotti-Robinson} (\emph{Kerr-BR}, in short) black holes. Let us now give its basic physical interpretation.

\subsection{Curvature singularities}
\label{sec:singularities}

It follows from \eqref{Psi2} that the \emph{curvature singularity} occurs at ${r=0}$, but \emph{only if} also ${\theta=\frac{\pi}{2}}$. It thus has a \emph{ring structure}, similarly as in the Kerr spacetime. When ${\theta\ne\frac{\pi}{2}}$ it is possible to reach the region ${r<0}$. At this curvature singularity also the electromagnetic field diverges.

\subsection{Horizons}
\label{sec:horizons}

The function \eqref{math-Q} enables us to easily determine the position of the horizons. They are located at ${{Q}=0}$ where the coordinate~$r$ changes its spatial/temporal character. The corresponding quadratic equation ${\Delta=0}$ given by \eqref{Delta} has up to two roots
\begin{align}
r_\pm= \dfrac{m\,I_2 \pm \sqrt{m^2I_2-a^2I_1^2}}{I_1^2-B^2m^2I_2} \,I_1 \,,  \label{rb}
\end{align}
which localize the \emph{outer} and \emph{inner} black hole horizons.

For ${B=0}$ implying ${I_1=1=I_2}$ we recover the formula ${r_\pm=m\pm\sqrt{m^2-a^2} }$ valid for the Kerr black hole.

In the case ${a=0}$ without rotation (so that ${I_1=1=I_2}$) we get a \emph{single} horizon at
\begin{align}
r_h= \dfrac{2m}{1-B^2m^2}  \,.  \label{rb-for-a=0}
\end{align}
Notice that this is located \emph{at greater values} than the Schwarzschild horizon ${r_{\rm Schw}=2m<r_h}$.

In the \emph{extreme} situation, which appears for the large value of the rotation~$a$ when ${a^2I_1^2=m^2I_2}$, the magnetic field~$B_{\rm extr}$ takes the extremal value
\begin{align}
B^2_{\rm extr}  = \frac{2}{a^4}\big(m-\sqrt{m^2-a^2}\,\big)\sqrt{m^2-a^2}\,.  \label{B-extr}
\end{align}
In such a case the two horizons coincide at
\begin{align}
r_{\rm extr} =  \frac{m}{I_1}  =  \frac{a^2}{m-\sqrt{m^2-a^2}}\,.  \label{rextr}
\end{align}
For ${a=m}$ we get ${B_{\rm extr}=0}$ and ${r_{\rm extr} = m}$ which is the extreme Kerr subcase.

\subsection{Ergoregions}
\label{sec:ergoregions}

For the black hole metric \eqref{Kerr-BR} the condition ${g_{tt} = 0}$ defines the boundary of the ergoregions, that are the surface of infinite redshift where also observers at fixed $r,\theta$ cannot ``stand still''. Such a boundary is located at
\begin{equation}
 Q(r_e) = a^2\sin^2\theta  \,P(\theta) \,,
\label{gtt=0}
\end{equation}
where the functions $P$, $Q$ are given by \eqref{tilde_P}, \eqref{math-Q}. For ${a=0}$ it coincides with a horizon given by ${Q=0}$. For ${a\not=0}$ the ergoregion boundary ``touchess'' the horizon at the poles ${\theta=0}$ and ${\theta=\pi}$. An explicit form of $r_e$ as a function of $\theta$ can be  found numerically and plotted, as in Fig.~\ref{Fig} where it is indicated by dashed lines.

\subsection{Regularity of the axes}
\label{sec:strings}

The spatial axes of symmetry are identified by zeros of the Killing vector field $\partial_\varphi$ norm. These are located at ${\theta=0}$ and ${\theta=\pi}$, and can be made \emph{simultaneously regular} by a proper choice of the unique \emph{conicity parameter~$C$} determining the range of the angular coordinate
\begin{equation}\label{range-of-C}
\varphi\in[0,2\pi C)\,.
\end{equation}
A circle around ${\theta=0}$ and ${\theta=\pi}$ given by ${\theta=\hbox{const.}}$, assuming fixed $t$ and~$r$, has an invariant length of its circumference ${\int_0^{2\pi C}\!\! \sqrt{g_{\varphi\varphi}}\, \dd\varphi}$, and  a radius ${\int\!\sqrt{g_{\theta\theta}}\, \dd\theta}$. Thus
\begin{equation}
\lim_{\theta\to0} \frac{\hbox{circ.}}{\hbox{rad.}}    = 2\pi C\,{P}(0),
 \qquad
\lim_{\theta\to \pi} \frac{\hbox{circ.}}{\hbox{rad.}} = 2\pi C\,{P}(\pi).
 \label{con0-conpi}
\end{equation}
The axis is regular if the fraction is equal to ${2\pi}$. Moreover, for $P$ of the form \eqref{tilde_P} we get ${P(0) = P(\pi) }$, so that there exists a unique choice of the conicity parameter
\begin{align}
 C = \Big[1 + B^2 \Big(m^2\,\dfrac{I_2}{I_1^2} - a^2 \Big)\Big]^{-1},
 \label{C}
\end{align}
simultaneously regularizing both the axes. For the Kerr black hole (${B=0}$) we get ${C=1}$, for the non-rotating Schwarzschild-BR black hole  (${a=0}$) we must choose~${ C = 1/(1 + B^2 m^2)}$.

\subsection{The electromagnetic field}
\label{sec:plotsEMfield}

All components of the Maxwell field are given by the scalars \eqref{Phi0,1=0,e=0=g}, \eqref{Phi1,l=0,e=0=g}. To clarify the character of such electromagnetic field in more detail, we determine the corresponding 3-vectors of the electric and magnetic fields $E^i$ and $B^i$, and visualize them in pictures.

The vectors $E^i$ and $B^i$ arise from the 1+3 splitting of a spacetime adapted to an observer with 4-velocity $\mathbf{u}$  and the transverse 3-dim local Cartesian frame ${\mathbf{e}^{(i)}=\mathbf{e}_{(i)}}$,
\begin{align}\label{EandB}
    E^i \equiv F^{\mu\nu}\,u_\mu e_\nu^{(i)}\,,\qquad
    B^i \equiv \tfrac{1}{2}\epsilon^{\mu\nu\rho\sigma}F_{\rho\sigma}\,u_\mu e_\nu^{(i)}\,,
\end{align}
where $F_{\mu\nu}$ are components of the 2-form ${\mathbf{F}=\dd \mathbf{A}^{\rm real}}$ given by \eqref{A-7}. For the metric \eqref{Kerr-BR} we chose the ZAMO observer 4-velocity,
\begin{align}\label{ZAMO}
    \mathbf{u} &= N^{-1}\big(\partial_t+\omega\,\partial_\varphi\big)\,,\quad\hbox{with}\nonumber\\
     N^2 &= \frac{PQ\rho^2}{R\,\Omega^2}\,,\quad
  \omega = \frac{a}{R}\big(P(r^2+a^2)-Q\big)\,,
\end{align}
${R=P(r^2+a^2)^2-Q\,a^2\sin^2\theta}$, and the frame adapted to $r, \theta, \varphi$ as
${\mathbf{e}_{(r)}=(\mathbf{k}-\mathbf{l})/\sqrt{2}}$, ${\mathbf{e}_{(\theta)}=(\mathbf{m}-\bar{\mathbf{m}})/(\sqrt{2}\,\im)}$, ${\mathbf{e}_{(\varphi)}=(\mathbf{m}+\bar{\mathbf{m}})/\sqrt{2}}$ where  ${\mathbf{k}, \mathbf{l}, \mathbf{m}}$ are given by \eqref{null_tetr_trans}.
Due to the axial symmetry ${B^{(\varphi)}=0}$, and ${B^{(r)}, B^{(\theta)}}$ can be plotted in 2-dim pictures in which $(r, \theta)$ are treated as the polar coordinates in the auxiliary $(x, y)$-plane, that is ${x=r\cos\theta}$, ${y=r\sin\theta}$, ${B^{(x)}= B^{(r)}\sin\theta + B^{(\theta)}\cos\theta}$, ${B^{(y)}= B^{(r)}\cos\theta - B^{(\theta)}\sin\theta}$.

\begin{figure}[t!]
\centerline{\includegraphics[scale=0.42]{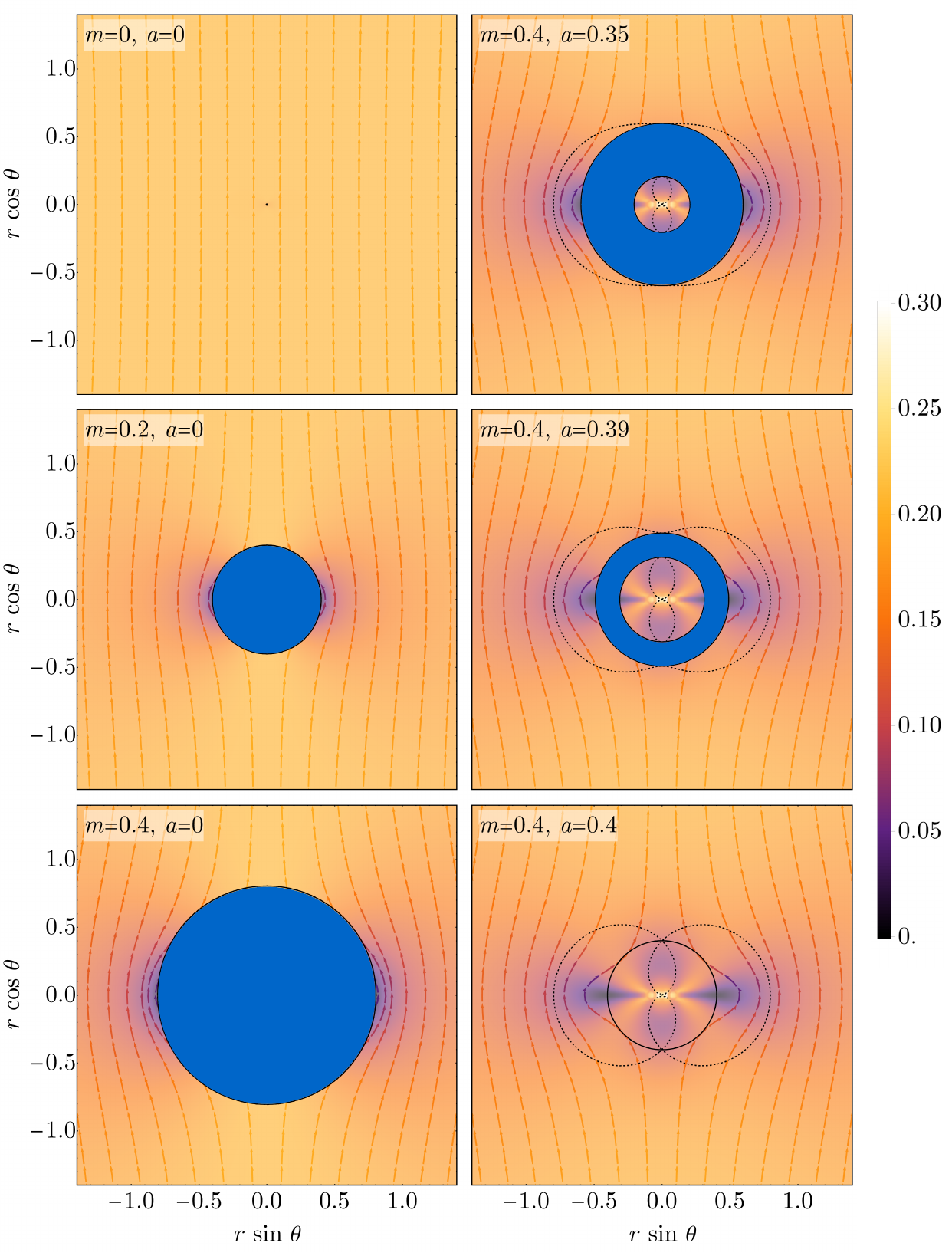}}
\vspace{0mm}
\caption{Visualization of the magnetic field by the color plots (encoding the magnitude of the field) and the force lines (the arrows indicating the directions of the field) for ${B=0.2}$. The top left panel (${m=0}$, ${a=0}$) shows the uniform magnetic field of the Bertotti-Robinson (BR) background universe. Left column corresponds to the Schwarzschild-BR black holes without rotation (${a=0}$), while the right column depicts the general situation for Kerr-BR black holes (${a\not=0}$) with the rotation parameters~${a=0.35, 0.39, 0.399998 }$ (labeled as ${a=0.4}$) and~${m=0.4}$. The horizons are indicated by the black circles, the ergoregions are bounded by the dashed curves. In all the cases the external magnetic field is weakened in the equatorial plane ${\theta=\frac{\pi}{2}}$, and the force lines are ``expelled away'' --- exhibiting the Meissner effect. The blue discs and annuli indicate non-stationary region ${Q<0}$ inside the black hole where the ZAMO observer \eqref{ZAMO} is not timelike and thus the field \eqref{EandB} is not well defined.
}
\label{Fig}
\end{figure}

In Fig.~\ref{Fig} we plot the magnetic field for several values of the physical parameters $m$ and $a$ (representing the mass and the rotation of the black hole) assuming the value ${B=0.2}$ and ${\gamma=0}$. (The pictures for ${\gamma=\frac{\pi}{2}}$ depicting the dual electric field are the same.) In these pictures the \emph{magnitude} of the magnetic field is represented by the \emph{color}, such that black corresponds to zero value, while the lighter colors (yellow to white) indicate bigger values. The specific \emph{direction} of the magnetic field in various points of the space is represented by the \emph{arrow} (which has everywhere the same size). Effectively, these arrows form the ``lines of forces'' visualizing the magnetic field. Horizons of the black holes are depicted by black circles, the ergoregions boundaries are indicated by dashed curves.

The top left panel shows the magnetic field in the spacetime with ${m=0}$, ${a=0}$. This is the uniform magnetic field of the Bertotti-Robinson (BR) universe. The middle left panel exhibits the situation in the \emph{Schwarzschild-BR} spacetime for ${m=0.2}$  (and ${a=0}$). It can be observed (both by the ``darker color'' near the equatorial plane ${\theta=\frac{\pi}{2}}$, and also by the ``deflection'' of the force lines) that the magnetic field is ``expelled'' from the black hole, away from its horizon. Such an effect --- called the \emph{Meissner effect} --- has already been reported in other axisymmetric spacetimes with black holes. In our case the expulsion is not complete because the black hole is not charged and extremal (as opposed to the Kerr-Newman-Melvin spacetime \cite{Bicak2015}). The bottom left  panel shows the same situation, but amplified by the larger Schwarzschild-BR black hole with bigger mass ${m=0.4}$ (and ${a=0}$).


The three panels on the right show the character of the magnetic field in the case when the black hole is rotating, that is in the \emph{Kerr-BR} spacetime for the same mass ${m=0.4}$. The value of the rotation parameter increases from ${a=0.35}$ to ${a=0.39}$, and to the extreme case ${a=0.4}$ (actually ${a\approx 0.399998}$). In all these cases the magnetic field is expelled away, forming more complicated patterns. The Meissner behavior, resulting from the non-linear interaction between the external (asymptotically) uniform Bertotti-Robinson magnetic field and the rotating  Kerr black hole is quite involved.

\subsection{Geodesic motion and ISCO}
\label{sec:geodesics}

Geodesics can be studied using integrals of motion related to symmetries of the spacetime \eqref{Kerr-BR}. The Killing vectors $\partial_t$ and $\partial_\varphi$ imply the conservation of energy and angular momentum of a uncharged test particle as
\begin{align}
    p_{t}=-E\,,\qquad p_{\varphi}=L\,,
\end{align}
respectively. A complete analysis of the geodesics exceeds the scope of this letter. To illustrate the effect of the external magnetic field, let us restrict here to the equatorial motion (${\theta=\frac{\pi}{2}}$) of a particle of a rest mass~$m_0$ in the non-rotating Schwarzschild-BR metric \eqref{metric-Schw-BR}.

\begin{figure}[ht!]
\centerline{\includegraphics[scale=0.131]{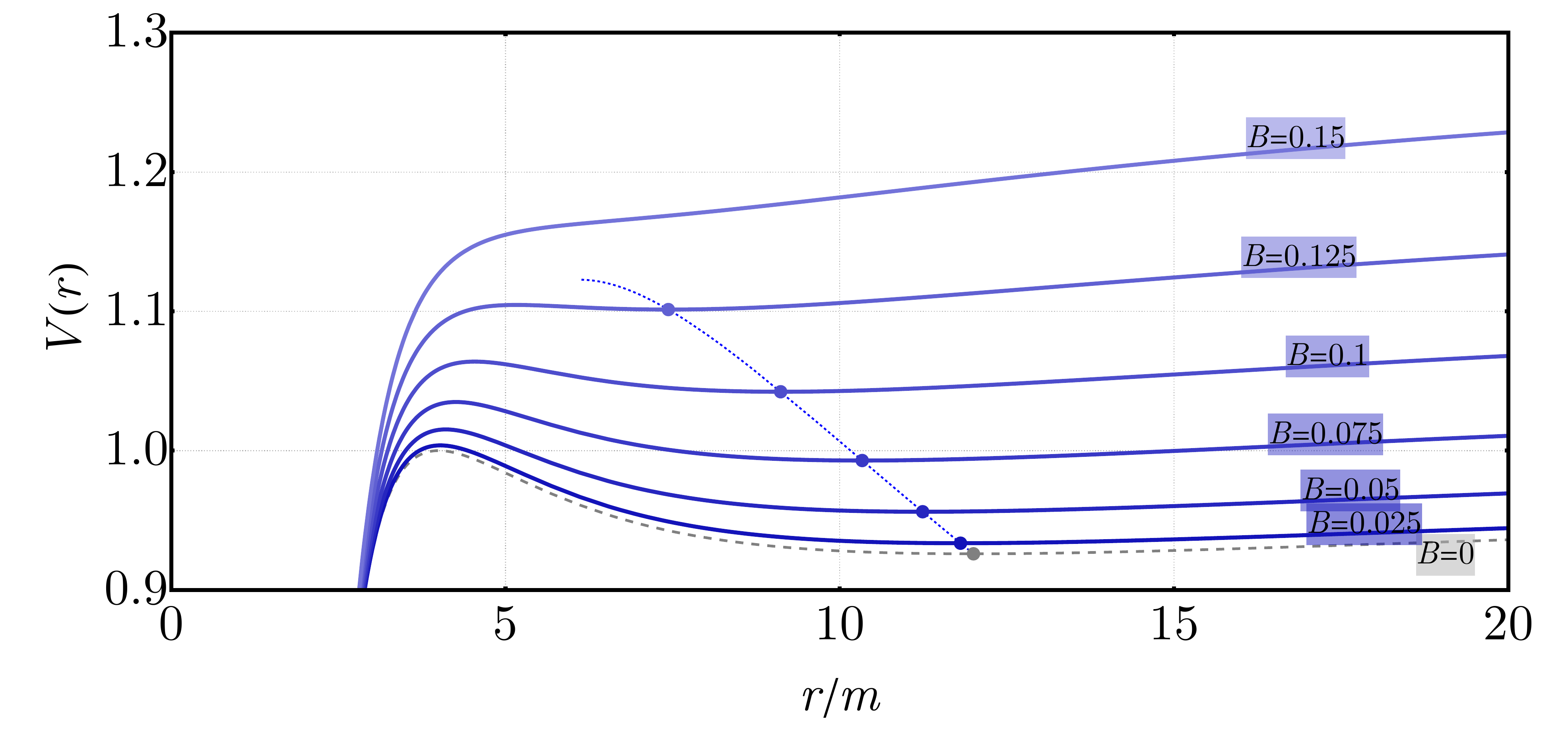}}
\vspace{0mm}
\caption{ Effective potential $V(r)$ as a function of the radial coordinate $r$ for different values of the magnetic field $B$ (for ${m=1}$, ${L=4m_0\ne L_{\rm ISCO}}$). The points on the dashed curve show the positions of the corresponding stable circular orbits.}
\label{Fig2}
\end{figure}

Using the normalization ${g_{\mu\nu}u^{\mu}u^{\nu}=-1}$ we get
\begin{align}
    \dot r^2=\Omega^4\Big[\,E^2/m_0^2-V(r)\Big].
\end{align}
Stable circular orbits (${\dot r=0}$) are possible only at the minimum of the effective potential (see Fig.~\ref{Fig2})
\begin{align}
\hspace{-3mm}
    V(r)=\Big( 1-B^2m^2-\dfrac{2m}{r}\Big)\Big[1+\dfrac{L^2}{m_0^2}\Big(\dfrac{1}{r^2}+B^2\Big)\Big].
    \label{V_expr}
\end{align}
From the conditions ${V'=0}$ and ${V''=0}$ we derive that the \emph{innermost stable circular orbit} (ISCO) is located at
\begin{align}
    r_{\rm ISCO}&=\dfrac{L^2_{\rm ISCO}(1-B^2m^2)}{2m(m_0^2+L^2_{\rm ISCO}B^2)},
   \quad \hbox{for}\nonumber \\
    L^2_{\rm ISCO}&=\dfrac{12m^2m_0^2}{(1-B^2m^2)^2-12B^2 m^2}\,.
\end{align}
Surprisingly, in view of \eqref{rb-for-a=0} we get a simple result
\begin{align}\label{rISCO}
    r_{\rm ISCO} = 3 r_h\,.
\end{align}
This is formally the same expression for the ISCO as in the Schwarzschild metric. Notice however, that the position of the horizon is not simply given by ${r_h = 2 m}$, but is modified by the presence of the magnetic field to
${r_h= 2m/(1-B^2m^2)>2m}$. The ISCO is thus located  \emph{at larger values} than in the Schwarzschild black hole.

\vspace{0mm}


\subsection{Thermodynamics}
\label{sec:thermodynamics}

Finally, we evaluate the basic thermodynamic quantities of this class of black holes, namely \emph{entropy}~$S$ and \emph{temperature}~$T$ of their horizons. Using the famous relations, see e.g. \cite{Wald:book1984}, these are respectively given by the horizon area~${\cal A}$ and the surface gravity~$\kappa$ as
\begin{equation}
S =\tfrac{1}{4}\,{\cal A}\,,
\qquad
T = \tfrac{1}{2\pi}\,\kappa\,.
\label{entropy and temperature}
\end{equation}

The \emph{horizon area} is obtained by integrating the angular coordinates of the metric \eqref{Kerr-BR} for fixed values of $t$ and ${r=r_h}$, namely ${\mathcal{A}(r_h) = \int_0^{2\pi C}\!\int_0^\pi \sqrt{g_{\theta \theta}\, g_{\varphi \varphi}}\,\,\dd \theta \, \dd \varphi}$, where $C$ is the conicity parameter \eqref{C}. Using  ${{Q}(r_h)=0}$ we get
${\mathcal{A} = 2\pi C\,\big(r_h^2+a^2\big) \int_0^\pi \Omega^{-2}(r_h)\,\sin\theta\,\dd \theta}$. Surprisingly, \eqref{Omega} evaluated on the horizon is a \emph{constant} ${\Omega^2(r_h) = 1+B^2r_h^2}$. By substitution ${x=\cos\theta}$ we obtain
\begin{equation}
\mathcal{A}_h = 4\pi C\,\frac{r_h^2+a^2}{1+B^2r_h^2 } \,.
\label{Ah}
\end{equation}
For \emph{non-rotating} black holes (${a=0}$) it reduces to ${\mathcal{A}_h = 4\pi C\,r_h^2/(1+B^2 r_h^2) = 16\pi \,m^2/(1+B^2 m^2)^3}$.

The \emph{surface gravity}~$\kappa$ of the horizon is defined as the ``acceleration'' of the null normal~$\xi^a$ generating the horizon at~$r_h$ via the relation ${\xi_{a;b}\,\xi^b  = \kappa\, \xi_a}$ (so that ${\kappa^2=-\frac{1}{2}\xi_{a;b}\,\xi^{a;b}}$) \cite{Wald:book1984}. In our case, this can be calculated using the usual formula ${\kappa_h = \tfrac{1}{2} {Q}'(r_h)/(r_h^2+a^2)}$. For \eqref{math-Q}, applying ${{Q}(r_h)=0}$, we get
\begin{align}
\kappa_{h}&= \dfrac{1+B^2r_h^2}{{r_{h}^2+a^2}}\,
    \Big( m\,\dfrac{I_2}{I_1} -\dfrac{a^2}{r_h}\Big )\,. \label{kappa-h}
\end{align}
In the Schwarzschild-BR black hole case (${a=0}$), the surface gravity of the black hole horizon of the metric \eqref{metric-Schw-BR} reduces to ${ \kappa_{h}= m\,(r_h^{-2}+B^2) = (1+B^2 m^2)^2/4m}$. Also, ${\kappa_{h}=0}$ for the \emph{extreme black hole} with a double-degenerate horizon at ${r_{\rm extr} =  m/I_1}$ which occurs when ${a^2I_1^2=m^2I_2}$. Such black holes have \emph{zero temperature}~$T$.

By combining \eqref{kappa-h}, \eqref{Ah} we obtain a nice relation
\begin{equation}
2\,T S \equiv \frac{1}{4\pi}\,\kappa_{h}\,\mathcal{A}_h
  =  C \Big( m\,\dfrac{I_2}{I_1} - \dfrac{a^2}{r_h} \,\Big).
\label{Smarr}
\end{equation}
It reduces to ${\,2\,T S = C m\,}$ when ${a=0}$. This resembles the \emph{Smarr formula} \cite{Smarr:1973}, namely ${M = 2\, T S}$, if we make the identification ${M\equiv C m}$. Moreover, it satisfies the first law of thermodynamics ${\delta M=T\,\delta S}$
if one assumes in its derivation that the conicity parameter is not changed (${\delta C=0}$). Similar assumption was used for the \hbox{C-metric} in~\cite{Appels2016}. However, alternative approaches exist, allowing the variations of $m$, $B$, and $C$ such that the relation \eqref{C} still holds. This may add terms related to the magnetic moment of a black hole (namely ${+\mu \,B}$ in the Smarr formula, and ${-\mu\,\delta B}$ in the first law), but such an analysis goes beyond the scope of the present work.

\section{Conclusions}

We presented the new family of spacetimes \eqref{Kerr-BR}, demonstrating that they describe rotating black holes in asymptotically uniform magnetic (or electric) external field. These are type~D exact solutions to the Einstein-Maxwell equations with a non-aligned electromagnetic field thus distinct from the Pleba\'nski-Demia\'nski class). The physical parameters are the black hole mass $m$, its rotation $a$, and the value of the field $B$. When they are set to zero, standard forms of the Kerr, Schwarzschild, and Bertotti-Robinson spacetimes are obtained.

The metric functions \eqref{rho2}--\eqref{Delta} are quadratic in $r$ and $\cos\theta$. Key physical characteristics of the new black holes can thus be easily evaluated, namely the location of singularities, horizons, ergoregions, regularity of the axis, geodesics, and thermodynamics. They generalize the Kerr black hole ones, with an interesting influence of the magnetic field. In particular, the Schwarzschild-BR (${a=0}$) black hole horizon ${r_h}$  \eqref{rb-for-a=0}, and the innermost stable circular orbit \eqref{rISCO} ${r_{\rm ISCO} = 3 r_h}$, are \emph{larger} than in the Schwarzschild case. With the magnetic field $B$, the black hole entropy~$S$ is \emph{smaller}, while its temperature~$T$ is \emph{bigger}. Interestingly, the Smarr formula \eqref{Smarr} holds, and the first law of thermodynamics is also~satisfied.

In Fig.~\ref{Fig} we visualized the magnitude and orientation of the magnetic field, given by \eqref{A-7} with respect to ZAMO observer \eqref{ZAMO}, for several values of $m$ and $a$. Both for the Schwarzschild-BR and the  Kerr-BR black holes it demonstrates the Meissner effect, namely that the external field is weakened in the equatorial plane, and its force lines are expelled away. This complements previous studies of such an effect in the context of (extreme) charged black holes \cite{Bicak2015, GurlebeckScholtz:2018}.

Black holes immersed in a magnetic field have already been studied in a number of works. These were mostly modeled by the Schwarzschild-Melvin spacetime and its generalizations to include rotation, charges, NUT, and other physical parameters \cite{Ernst1976_2}--\cite{Pravda2005}. It should be emphasized that such spacetimes are of algebraic type I.

Our novel solution is of type D, and the external Maxwell field is uniform without the black hole. Generalization to include charge and acceleration is possible, retaining the type~D structure \cite{OP-prepar2}.

We hope that the simple new metric will become useful for investigation of various effects, both in the context of mathematical theory of black holes, and also as a model of more realistic black holes in relativistic astrophysics.

\section*{Acknowledgments}

This work has been supported by the Czech Science Foundation Grant No.~GA\v{C}R 23-05914S and by the Charles University Grant No.~GAUK 260325. The authors thank David Kubiz\v{n}\'ak and Finnian Gray for valuable discussion concerning the thermodynamics, and Marcello Ortaggio for useful comments on the static case.

\end{document}